# A Physics-Driven Neural Network with Parameter Embedding for Generating Quantitative MR Maps from Weighted Images

**Lingjing Chen (1 and 2), Chengxiu Zhang (1 and 2), Yinqiao Yi (1 and 2), Yida Wang (1 and 2), Yang Song (3), Xu Yan (3), Shengfang Xu (4), Dalin Zhu (4), Mengqiu Cao (3), Yan Zhou (5), Chenglong Wang (1 and 2), Guang Yang (1 and 2)**

((1) Shanghai Key Laboratory of Magnetic Resonance, School of Physics and Electronic Science, East China Normal University, Shanghai, China, (2) Institute of Magnetic Resonance and Molecular Imaging in Medicine, East China Normal University, Shanghai, China, (3) MR Research Collaboration Team, Siemens Healthineers, Shanghai, China, (4) Department of Radiology, Gansu Provincial Maternity and Child-care Hospital, Lanzhou, China, (5) Department of Radiology, Renji Hospital, School of Medicine, Shanghai Jiao Tong University, Shanghai, China)

**Abstract:**

Purpose: To develop a deep learning-based approach that integrates MRI sequence parameters to improve the accuracy and generalizability of quantitative image synthesis from clinical weighted MRI.

Methods: We proposed a physics-driven neural network that embeds MRI sequence parameters — repetition time (TR), echo time (TE), and inversion time (TI) — directly into the model via parameter embedding. This design enables the network to learn the underlying physical principles of MRI signal formation. The model takes conventional $T_1$-weighted, $T_2$-weighted, and $T_2$-FLAIR images as input and synthesizes $T_1$, $T_2$, and proton density (PD) quantitative maps. The model was trained on healthy brain MR images and evaluated on both internal and external test datasets.

Results: The proposed method achieved high performance with PSNR values exceeding 34 dB and SSIM values above 0.92 for all synthesized parameter maps. It outperformed conventional deep learning models in accuracy and robustness, including data with previously unseen brain structures and lesions. Notably, our model accurately synthesized quantitative maps for these unseen pathological regions, highlighting its superior generalization capability. Conclusion: Incorporating MRI sequence parameters via parameter embedding allows the neural network to better learn the physical characteristics of MR signals, significantly enhancing the performance and reliability of quantitative MRI synthesis. This method shows great potential for accelerating qMRI and improving its clinical utility.

# 1. Introduction

Quantitative MRI (qMRI) creates parametric maps that directly measure tissue properties, such as $T_1$, $T_2$, and proton density (PD). These maps provide more accurate and stable information compared to traditional MRI images. By enabling the direct assessment of tissue microstructure and biophysical properties, qMRI enhances our ability to investigate the mechanisms underlying various diseases and offers better tissue characterization in both healthy and pathological states[1]. This capability makes qMRI an essential tool for diagnosing and monitoring pathologies like brain tumors[2], multiple sclerosis[3], and neurodegenerative diseases[4][5]. Furthermore, recent studies have widely used mean values of quantitative parameters extracted from regions of interest (ROIs) on qMRI maps as quantitative biomarkers for various diseases, including Alzheimer's disease[6] and neurodevelopmental disorders[7]. By providing more accurate quantitative information about the pathological changes, qMRI may facilitate a better understanding of these complex diseases. Additionally, parametric maps generated by qMRI are more robust than conventional weighted images, as they are less influenced by variations in MRI hardware and sequence parameters[8]. This enhances objectivity and reproducibility of image analysis and facilitates reliable comparisons across subjects, sites, and time points, making qMRI especially valuable for longitudinal studies and multi-center research[9-12].

However, traditional qMRI techniques face several challenges, particularly with regard to long scan time and low signal-to-noise ratios (SNR). Conventional qMRI methods typically require multiple acquisitions of the same anatomical region with varying pulse sequence parameters, such as repetition time (TR), echo time (TE), inversion time (TI). These acquisitions are followed by voxel-wise nonlinear fitting to generate quantitative parameter maps. This process is not only time-intensive but also requires extended patient immobility, which can lead to discomfort and reduced compliance. Consequently, the extended scan time associated with traditional qMRI techniques posed a substantial barrier to their widespread adoption in clinical practice.

Recently, simultaneous multiple-quantitative MRI imaging techniques, such as multi-dynamic multi-echo (MDME) sequence[13], MR fingerprinting (MRF)[14] and multi-pathway multi-echo (MPME) imaging[15], and STrategically Acquired Gradient Echo (STAGE)[16][17] have brought about significant advancements to qMRI. These novel qMRI techniques enabled the efficient acquisition of quantitative parameter maps, offering improved time efficiency compared to traditional qMRI approaches. However, despite these advancements, several limitations remain. For example, the image quality of some of these new techniques may be more sensitive to patient motion, and some of them may fail to provide all common quantitative maps. Moreover, these techniques rely on specialized sequences that are often not yet integrated into standard clinical protocols[18][19][20]. This reliance on non-standard protocols reduces their availability in typical clinical settings and hinders their broader adoption.

In recent years, deep learning has emerged as a transformative and groundbreaking tool for synthesizing conventional magnetic resonance (MR) images. Methods based on deep learning have demonstrated the state-of-the-art performance and surpassed traditional techniques in terms of accuracy, efficiency, and robustness. As qMRI gains increasing attention in the biomedical and clinical domains, there has been a growing interest in leveraging deep learning to synthesize qMRI images[21-26]. Recent studies have demonstrated the potential of deep learning in synthesizing qMRI maps, allowing for faster data acquisition and improved image quality. Pei et al.[25] used U-Net to accelerate the qMRI imaging by reducing the number of required weighted images. In contrast, Sun et al.[24] and Qiu et al.[26] employed U-Net to synthesize qMRI directly from conventional qualitative MR images, effectively bypassing the need for extensive data acquisition. Besides U-Net, generative adversarial networks (GANs) have also been used for MRI image generation. For example, Dar et al.'s pGAN[27] leverages adversarial training to produce high-fidelity, multi-contrast MRI images in a single unified model. While the above approaches mainly used image prior knowledge or anatomy prior knowledge for image generation, physics prior knowledge was gradually adopted as a useful addition[28][29][30]. For example, Shaikh et al.[28] proposed a physics inspired deep learning model that learns a correction map to reduce contrast mismatches in synthetic MRI by implicitly accounting for unmodeled effects like diffusion and magnetization transfer. Meanwhile, both Qiu et al.[29] and Jacobs et al.[30] incorporated physics-based losses into the training process by simulating contrast-weighted images from predicted quantitative maps using Bloch equation models, allowing the network to learn with physically consistent supervision. However, the synthetic contrast-weighted images were generated under fixed scan parameters, limiting the model's adaptability to varying acquisition settings. For data acquired with different scan parameters, these models require additional retraining or fine-tuning.

Although deep learning has demonstrated notable success in qMRI synthesis, a significant common limitation of most existing methods is their failure to account for the varying scanning parameters of MRI acquisition protocols, which leads to contrast variations in weighted images scanned with the same pulse sequence. As a result, these methods are constrained to handling only input images scanned with fixed parameters, thereby lacking the flexibility required by applications involving diverse imaging conditions.

In this study, we proposed a physics-driven neural network with parameter-embedding to generate multiple parametric maps ($T_1$, $T_2$, and PD maps) from $T_1$-weighted ($T_1$w), $T_2$-weighted ($T_2$w) and $T_2$-fluid-attenuated inversion recovery ($T_2$-FLAIR) images. Different from previous works, our proposed model used both weighted images and the corresponding scan parameters as input, so that the model can be used to generate quantitative maps from weighted images scanned with different scan parameters. Our proposed model achieved remarkable performance on both internal and external datasets, even with unseen anatomical structures, which further validated its generalization capability.

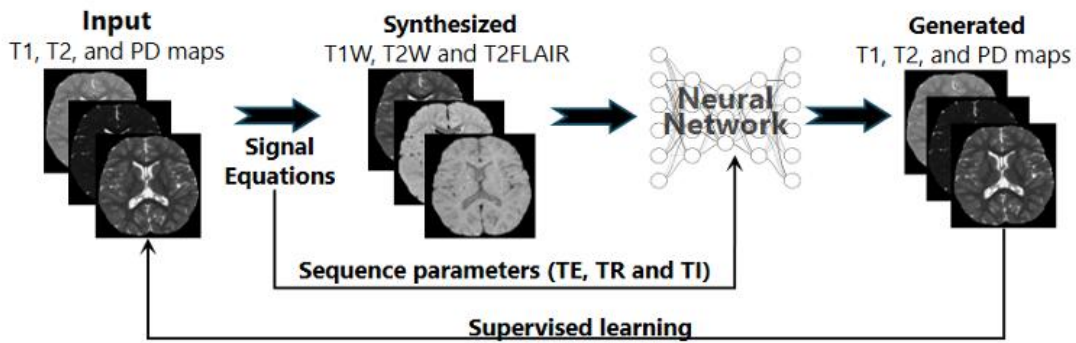

Figure 1. The workflow of this study.

## 2. Methods

The workflow of the study is shown in Figure 1. First, $T_1$w, $T_2$w and $T_2$-FLAIR images were synthesized from qMRI images acquired by MDME using signal formula for turbo spin echo (TSE) and FLAIR pulse sequences. Randomization of TE, TR and TI was employed to simulate varying scan protocols. Then, the synthesized weighted images together with the corresponding scan parameters were used to train a neural network which simultaneously generated $T_1$, $T_2$, and PD maps from $T_1$w, $T_2$w, and $T_2$-FLAIR images. Finally, the proposed model was evaluated with both internal and external test datasets.

### 2.1 Datasets

This study retrospectively enrolled a total of 119 subjects from two different centers. Since the research involved only anonymized medical records and did not require patient intervention, the Institutional Review Board (IRB) of the Ethics Committee of Gansu Provincial Maternity and Child-care Hospital and Ethics Committee of Renji Hospital granted an exemption from formal ethics approval and waived the requirement for informed consent. The first dataset included 69 children who were admitted to Gansu Provincial Maternity and Child-care Hospital (Center I) . Among them, 34 healthy infants constituted the training set, while the internal test set included 35 subjects diagnosed with congenital heart disease (CHD) or hepatic biliary (HB) disorders. 50 subjects were collected from Renji Hospital (Center II) including 47 healthy adults and 3 patients (meningioma, subdural arachnoid cyst and brain atrophy). Detailed inclusion criteria are in Supplementary material.

All subjects from Center I underwent MRI examinations on a 3T MRI scanner (MAGNETOM Vida, Siemens Healthineers, Germany). 69 subjects from Center I underwent Synthetic MRI examinations using a MDME sequence with the following scanning parameters: TR 4750 ms, TE 23 and 103 ms, receiver bandwidth 150 Hz/voxel, the field of view = 180 × 180 mm², matrix size = 256 × 180, 3 mm slice thickness, 3.9 mm slice gap, and a total scan time of 4 min 50s.

29 subjects from Center II underwent MR scans on a 3T MRI scanner (MAGNETOM Vida, Siemens Healthineers, Germany). Synthetic MRI utilized a MDME sequence with the following scanning parameters: TR 4350 ms, TE 23 and 103 ms, receiver bandwidth 310 Hz/voxel, the field of view = 180 × 180 mm², matrix size = 384 × 234, 5 mm slice thickness, 6.5 mm slice gap. 21 subjects from Center II underwent MR scans on a 3T MRI scanner (Ingenia, Philips Healthcare, Netherlands). Specifically, TSE protocol was performed with the following scanning parameters: $T_1$w with TR=229.69 ms and TE=2.3 ms; $T_2$w with TR=2364.65 ms and TE=90 ms; $T_2$-FLAIR with TR=7000 ms, TE=120 ms and TI=2250 ms, receiver bandwidth 150 Hz/voxel, the field of view = 192 × 192 mm², matrix size = 230 × 230 mm slice thickness, 6.5 mm slice gap. A summary of the dataset used in this study is shown in Table 1.

Table 1. Statistics of datasets used in this study.

| | Training Set | Internal Test Set | External Test Set 1 | External Test Set 2 |
|---|---|---|---|---|
| Subjects | 34 | 35 | 29 | 21 |
| Hospital | Gansu Provincial Maternity and Child-care Hospital | Gansu Provincial Maternity and Child-care Hospital | Renji Hospital | Renji Hospital |
| Scanner | Siemens 3.0T | Siemens 3.0T | Siemens 3.0T | Philips 3.0T |
| Sequence | MDME | MDME | MDME | TSE |
| Population Characteristics | Healthy children | Children with CHD and HB | 26 healthy adults and 3 patients | Healthy adults |

# CHD: congenital heart disease, HB: hyperbilirubinemia.

## 2.2 Weighted image generation

The weighted MR images can be analytically derived from parametric maps given required sequence parameters. For conventional sequences, established signal equations can be used to map tissue parameters ($T_1$, $T_2$, and PD) to MR intensities as functions of scan parameters (TE, TR, and TI). In more complicated cases, more sophisticated methods are needed[25]. In this study, we synthesized $T_1$w, $T_2$w, and $T_2$-FLAIR images from acquired $T_1$, $T_2$, and PD maps using signal equations for TSE and FLAIR sequences:

$$S_{TSE} = PD \left(1 - e^{\frac{-TR}{T_1}}\right) e^{-\frac{TE}{T_2}} \tag{1}$$

$$S_{FLAIR} = PD \left(1 - 2e^{-\frac{TI}{T_1}} + e^{-\frac{TR}{T_1}}\right) e^{-\frac{TE}{T_2}} \tag{2}$$

To generate images with various contrast, sequence parameters were randomized within specified ranges given in Supplementary Material Table S2. The ranges for randomization were set deliberately larger than those used in typical clinical settings in order to increase the diversity of the image contrast and maximize model generalizability. This intentional expansion of parameter range forced the model to learn the underlying physics to generate qMRI maps from weighted images acquired with diverse scan parameters, rather than depending too much on the image prior knowledge learnt from images acquired with a fixed scan protocol. As demonstrated in Figure 2, varying TE, TR, and TI values produced distinct contrast in synthesized weighted images while maintaining structural consistency.

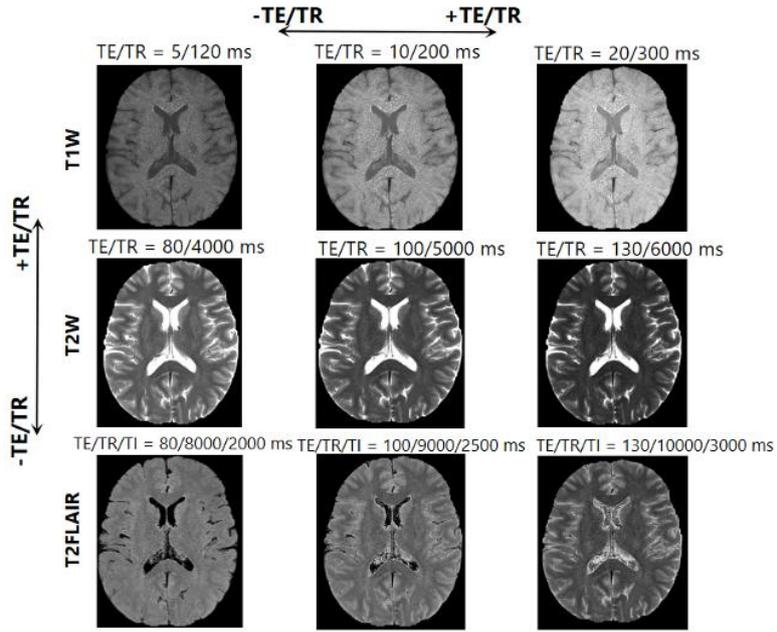

Figure 2. $T_1$w, $T_2$w and $T_2$-FLAIR images synthesized with different sequence parameters from one same set of the $T_1$, $T_2$, and PD maps.

## 2.3 Data preprocessing

Data preprocessing played a crucial role in enhancing the model's robustness and generalization capabilities. Inspired by the success of nnUNet[31] and tailored to the study's objectives, we implemented a standardized pipeline for data preprocessing. First, skull stripping was applied to all parametric maps using the Brain Extraction Tool (BET) [32] to remove non-brain tissues. For each sequence of each case, voxel values were clipped to the [0.5, 99.5] percentile range to mitigate outlier effects. Next, $T_1$w, $T_2$w, and $T_2$-FLAIR images were computed based on Eqs. (1)-(2). This was followed by modality-specific scaling: $T_1$w, $T_2$w, and $T_2$-FLAIR images were normalized by dividing by their respective maximum intensity values. Finally, all images were center-cropped to a matrix size of 256 × 256 pixels before used as input of the neural network.

## 2.4 Model structure

The architecture of the proposed Physics-Driven Parameter-Embedded Network (PDPE-Net) is shown in Figure 3. It adopts a U-Net like encoder-decoder structure with multiple input/multiple output capabilities. Three conventional weighted images ($T_1$w, $T_2$w, $T_2$-FLAIR) are fed into individual physical encoding branch as input to simultaneously generate three output channels representing $T_1$, $T_2$, and PD parametric maps. The feature maps extracted from the three weighted images are concatenated at the bottom layer. One core module of our physics-driven neural network is *the parameter embedding module*, where the scanning parameters are embedded into a feature matrix and passed through a normalization layer, which maps each parameter into a high-dimensional latent space. On each level of the encoding path, the parameter embeddings are resized to match the feature map dimensions and concatenated along the channel axis. The concatenated feature maps are then passed through two convolution layers with kernel size of $1 \times 1$. This operation, similar to a multi-layer perceptron network, enables voxel-wise computations for feature fusion. Each $1 \times 1$ convolution acts as a fully connected layer applied independently to each voxel, allowing the network to integrate complex feature interactions. This ensures that the feature maps of each modality incorporate relevant physical parameter information. More detailed network architecture is described in Supplementary material.

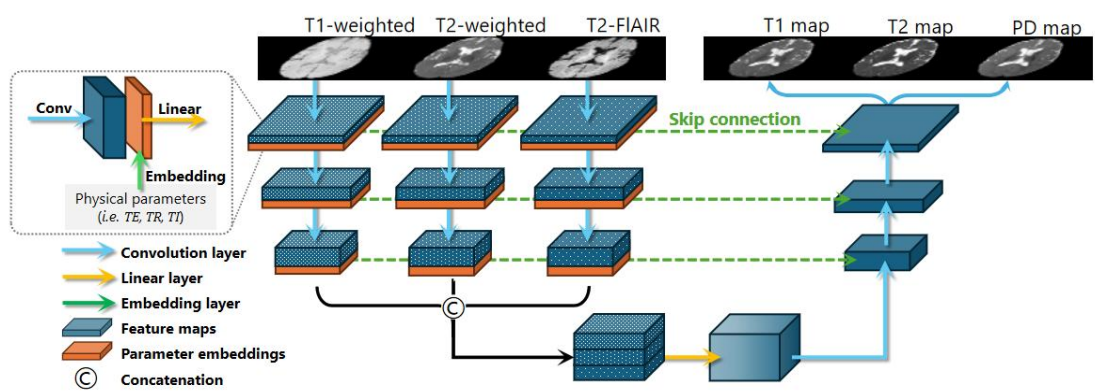

Figure 3. The architecture of PDPE-Net. Weighted images and associated scan parameters are used as input. Each weighted image follows its own downsampling path, with feature maps concatenated with associated scan parameters. Multi-scale features are fused at the lowest level before upsampled to generate final quantitative images. Inputs: $T_1$w, $T_2$w, and $T_2$-FLAIR slices; Target: $T_1$, $T_2$, and PD maps.

## 2.5 Model training

To enhance the generalizability of the model, online data augmentation was used during training, including: (1) random horizontal flipping in the axial plane, and (2) random scaling with factors between 0.9 and 1.1. These transformations introduced variability in image appearance, orientation, and scale while maintaining anatomical plausibility.

L2 loss of the brain region extracted by BET was used as loss function. Compared to L1 loss, L2 loss is more robust to artifacts, noise, and misregistration errors[33]. Total loss was the sum of loss for the three quantitative maps ($T_1$, $T_2$, and PD maps), denoted as:

$$L_{Total} = \lambda_1 L_{T1} + \lambda_2 L_{T2} + \lambda_3 L_{PD} \tag{3}$$

where the respective weighting factors, $\lambda_1$, $\lambda_2$, and $\lambda_3$ were set to 0.1, 0.5, and 0.15, respectively, to balance the loss magnitudes across different quantitative maps.

The model was implemented using PyTorch 2.0.1 and trained from scratch with the Adam optimizer[34], with an initial learning rate of 0.001. Training was conducted for up to 3000 epochs with early stopping triggered if no performance improvement was observed for 150 consecutive epochs, mitigating overfitting risks. The best model checkpoint was selected based on the sum of PSNR values for $T_1$, $T_2$, and PD maps on the validation set, and was saved when this cumulative metric reached a new maximum. The training was performed on a Linux workstation equipped with NVIDIA GeForce GTX 1080 Ti GPU.

## 2.6 Model evaluation

We used peak signal-to-noise ratio (PSNR), mean absolute error (MAE), mean percentage error (MPE) and structural similarity index (SSIM) to quantitatively evaluate metrics. These metrics are defined as follows:

$$MAE(x, y) = \frac{1}{N}\sum_{i=1}^{N} |x_i - y_i| \tag{4}$$

$$MPE(x, y) = \frac{1}{N}\sum_{i=1}^{N} \left|\frac{x_i - y_i}{y_i}\right| \tag{5}$$

$$PSNR(x, y) = 10 \log_{10}\left(\frac{\max(y)^2}{MSE(x, y)}\right) \tag{6}$$

where MSE(x, y) is the Mean Squared Error (MSE) between the two images, calculated as:

$$MSE(x, y) = \frac{1}{N}\sum_{i=1}^{N} (x_i - y_i)^2 \tag{7}$$

where $x_i$ and $y_i$ represents voxels in the brain parenchyma on the predicted map and ground-truth map, respectively. N is the number of voxels in the brain region. Details on brain parenchyma region extraction are in Supplementary material.

$$\text{SSIM}(x, y) = \frac{2\mu_x\mu_y+C_1}{\mu_x^2+\mu_y^2+C_1} \cdot \frac{2\sigma_{xy}+C_2}{\sigma_x^2+\sigma_y^2+C_2} \tag{8}$$

where $\mu_x$ and $\mu_y$ are the mean voxel values of local image patches x and y; $\sigma_x^2$ and $\sigma_y^2$ are the variance of x and y, respectively; $\sigma_{xy}$ is the covariance between x and y. In this study, each local patch is of size $11 \times 11$ pixels.

To evaluate the predicted quantitative values of brain tissue, we manually selected some regions of interest (ROIs) in different anatomical structures using ITK-SNAP (v4.0.1), as illustrated in Supplementary Material Figure S1. Each ROI was circular in shape with a diameter of 8 pixels and was placed in uniform regions of WM, GM and CSF to calculate mean $T_1$, $T_2$, and PD values of different brain tissues.

We compared the performance of PDPE-Net against the baseline model (U-Net without parameter embedding) and pGAN[27] on both the internal test set and external test set 1. Note that PD values have no absolute meaning, so they were normalized to the range of 0 to 1.

For visual assessment, we presented both the predicted and ground-truth parametric maps from the test set. To better visualize the differences between the images, we calculated normalized squared error (NSE) maps between the predicted and ground-truth $T_1$, $T_2$, and PD maps, denoted as:

$$\text{NSE}(x, y) = \frac{(\text{MAP}(x)-\text{MAP}(y))^2}{\text{MAP}(y)} \times 100\% \tag{9}$$

where MAP represents one of the $T_1$, $T_2$, or PD maps.

## 3. Results

### 3.1 Quantitative evaluation

Quantitative evaluation of our proposed PDPE-Net's performance on the internal test set and the external test set 1 is summarized in Table 2. Here we also present the comparative results of U-Net (without parameter embedding) and pGAN. Notably, PDPE-Net consistently demonstrated superior performance across both test sets. Specifically, PDPE-Net achieved a PSNR exceeding 34 dB, an SSIM above 0.92, MAE less than 0.02, and MPE less than 0.08 (P < 0.05), outperforming the baseline model. These results underscore PDPE-Net's capability to preserve structural similarity and enhance image fidelity by embedding sequence parameters.

Table 2. Comparison of the performance of PDPE-Net and two other models.

| | | Internal test set | | | External test set 1 | | |
|---|---|---|---|---|---|---|---|
| | | p-GAN | U-Net | PDPE-Net | p-GAN | U-Net | PDPE-Net |
| SSIM | $T_1$ | 0.9588±0.0088 | 0.9813±0.0064 | **0.9877±0.0036** | 0.9217±0.0288 | 0.9600±0.0099 | **0.9720±0.0112** |
| | $T_2$ | 0.9425±0.0212 | 0.9653±0.0135 | **0.9715±0.0075** | 0.9628±0.0067 | 0.9527±0.0103 | **0.9802±0.0053** |
| | PD | 0.9279±0.0217 | 0.9614±0.0184 | **0.9678±0.0107** | 0.8647±0.0606 | **0.9344±0.0211** | 0.9292±0.0225 |
| PSNR | $T_1$ | 27.4813±2.5820 | 31.2239±3.2741 | **34.9550±2.1031** | 26.5128±1.0198 | 27.9828±1.6427 | **34.3942±2.1815** |
| | $T_2$ | 34.6401±2.5550 | 40.5736±2.7983 | **41.9794±2.9087** | 38.7354±2.5345 | 40.1152±1.2864 | **44.1918±2.5000** |
| | PD | 29.4447±2.8058 | 34.7036±2.7372 | **37.4389±2.5052** | 28.6664±1.8106 | 33.3641±1.6622 | **34.0620±1.7561** |
| MAE | $T_1$ | 0.0304±0.0147 | 0.0217±0.0109 | **0.0099±0.0023** | 0.0396±0.0064 | 0.0345±0.0076 | **0.0149±0.0040** |
| | $T_2$ | 0.0077±0.0040 | 0.0054±0.0029 | **0.0031±0.0008** | 0.0065±0.0009 | 0.0071±0.0012 | **0.0031±0.0053** |
| | PD | 0.0245±0.0072 | 0.0141±0.0045 | **0.0091±0.0024** | 0.0281±0.0066 | 0.0164±0.0029 | **0.0155±0.0033** |
| MPE | $T_1$ | 0.0912±0.0301 | 0.0831±0.0340 | **0.0515±0.0326** | 0.2097±0.0395 | 0.1909±0.0409 | **0.0790±0.0292** |
| | $T_2$ | 0.1173±0.0451 | 0.0921±0.0341 | **0.0577±0.0105** | 0.1679±0.0341 | 0.1877±0.0381 | **0.0807±0.0152** |
| | PD | 0.0771±0.0082 | 0.0455±0.0076 | **0.0313±0.0076** | 0.1025±0.0229 | 0.0636±0.0313 | **0.0662±0.0606** |

\# Values are expressed in the form of mean ± std. The best scores are marked in bold.

Figure 4 illustrates the correlations between predicted and true mean $T_1$, $T_2$, and PD values in WM, GM and CSF regions for both internal test set and external test set 1. The majority of data points align closely with the diagonal, demonstrating accurate predictions.

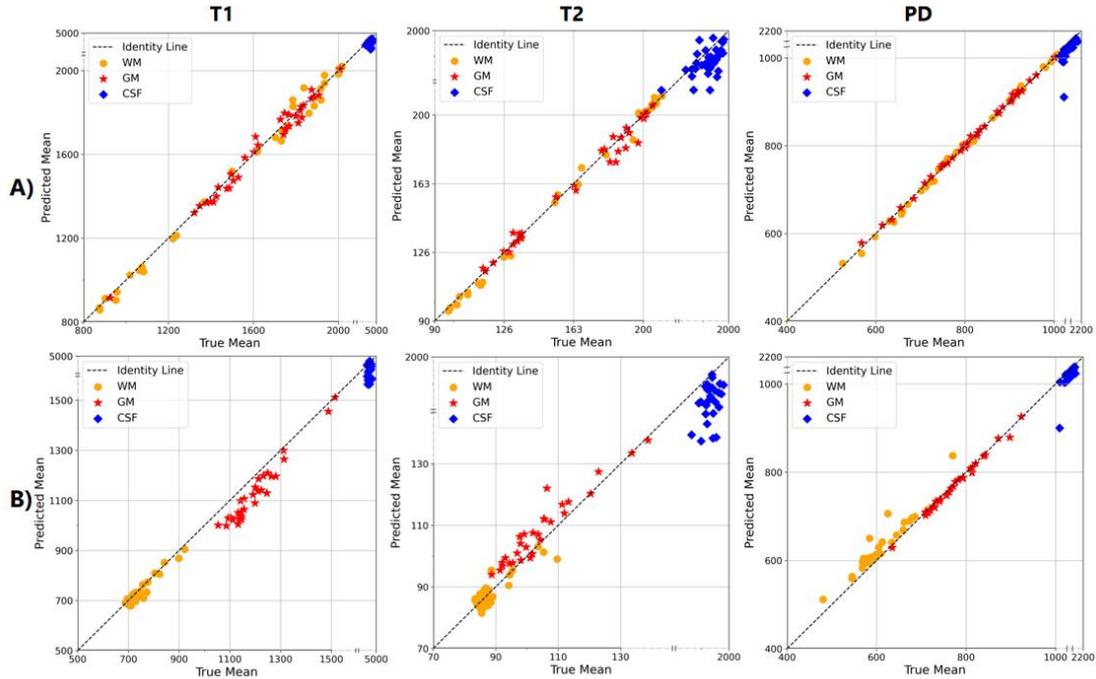

Figure 4. Correlation between the computed mean $T_1$, $T_2$ and PD values and the reference

values, calculated from manually delineated ROIs. Note that the x- and y-axes include a linear segment (solid lines) and a logarithmic segment (dashed lines). A) Internal test set B) external test set 1.

#Note: The double bars on the axes indicate a switch from linear to log10 scale.

**3.2 Visual evaluation**

Figure 5 presents a visual comparison of our proposed PDPE-Net with U-Net and pGAN using images in both the internal test set and external test set 1. From the NSE maps, the main differences between the predicted and ground-truth maps occur at the brain boundaries and the CSF region. It can be seen clearly that the proposed method generates results of the highest-quality across both datasets.

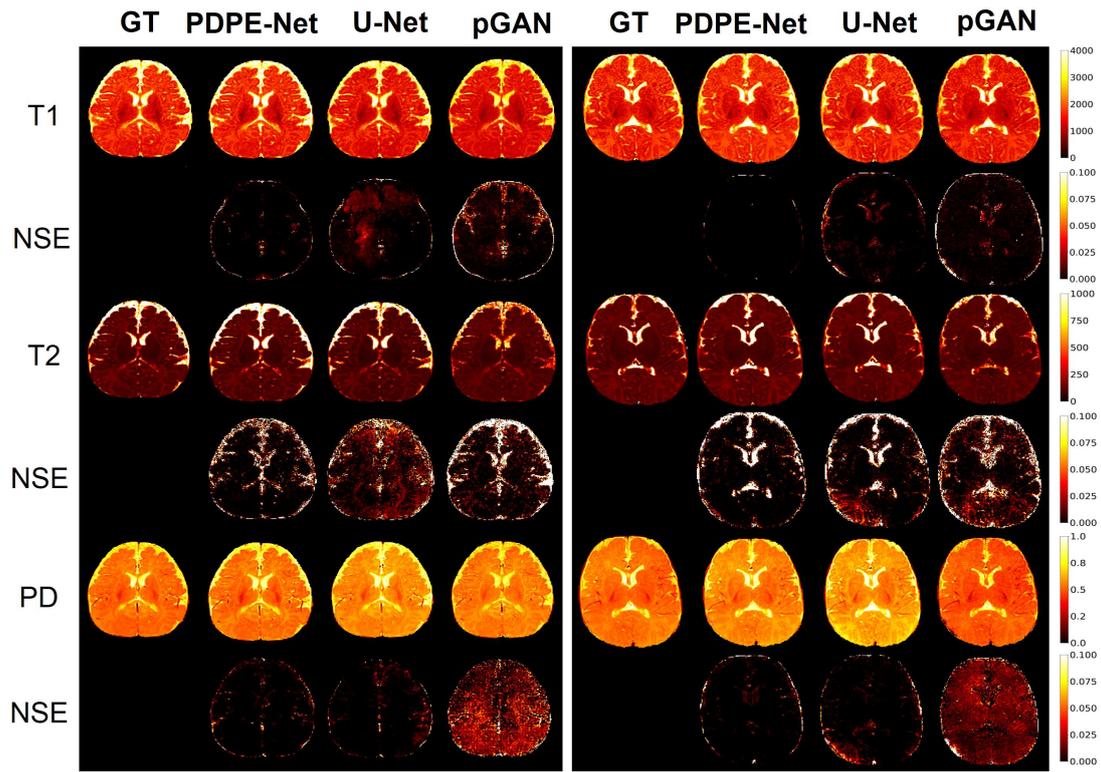

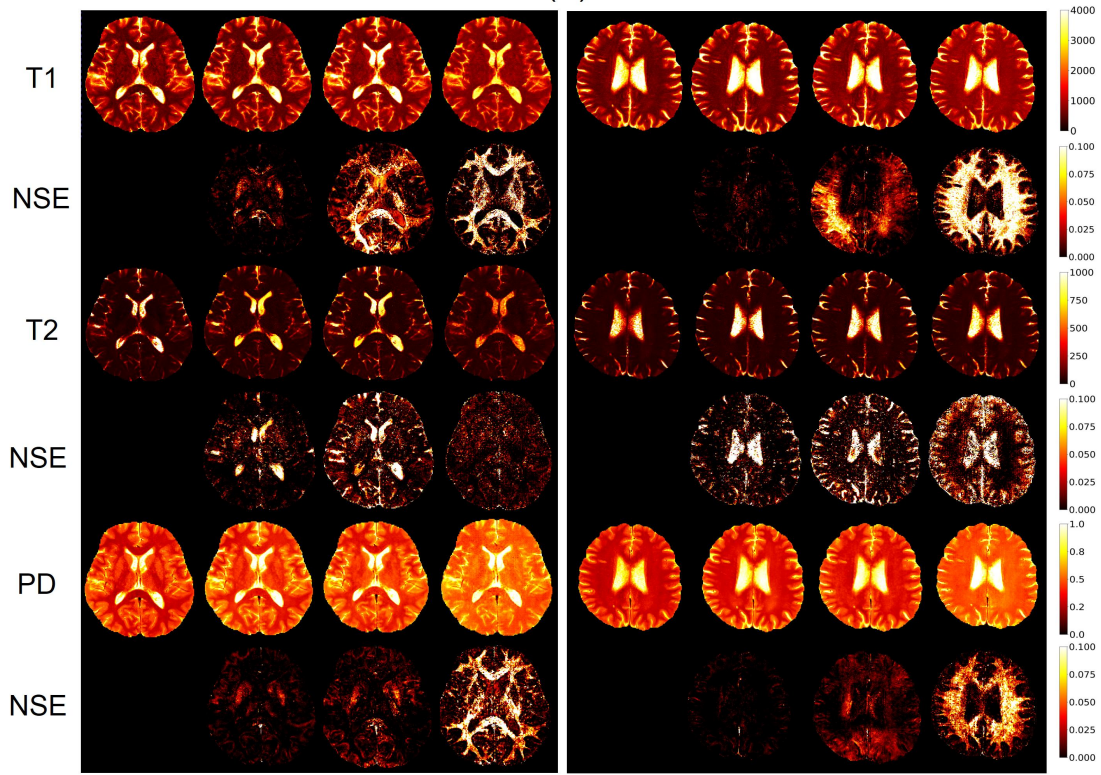

Figure 5. Comparison of $T_1$, $T_2$, and PD maps generated by different models. A) Two cases from internal test set. B) Two cases from external test set 1.

**3.3 Generalization capability evaluation**

Figure 6 presents slices from two patients in external test set 1. Although the model was exclusively trained on healthy infant brain data, it successfully predicted accurate quantitative values even when applied to previously unseen pathological brain tissue.

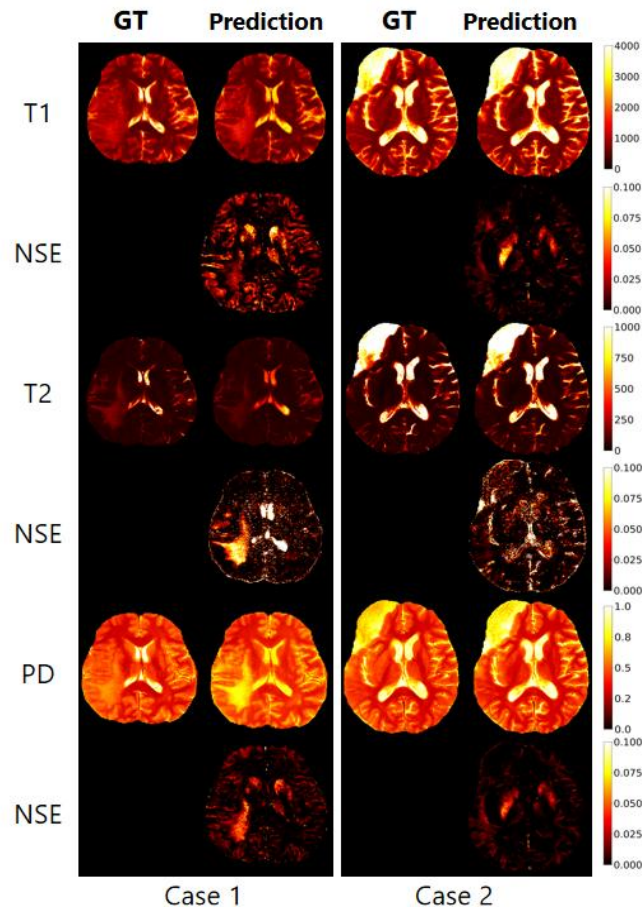

Figure 6. Slices from two patients in the external test set 1 are shown. GT refers to the ground truth corresponding to the slices.

Our model was also evaluated on real-world weighted images acquired from a different vendor. Figure 7 presents two representative axial slices of the $T_1$, $T_2$, and PD maps that were generated from weighted images using our trained model. The generated parametric maps exhibit visually realistic and effectively capture most of the structural details. Although ground truth quantitative maps were unavailable, we computed mean values for WM, GM, and CSF using ROIs defined in Supplementary Material Figure S1, as summarized in Table 3. These values align with normal range reported in the literature[35], further supporting the reliability of our model. Notably, predictions remained robust even when tested on images acquired with parameter values beyond the training range, demonstrating strong generalization capability.

Table 3. Mean intensity values of $T_1$, $T_2$, and PD within WM, GM, and CSF ROIs in External Test Set 2.

|   | WM | GM | CSF |
|---|---|---|---|
| $T_1$ | 789.78±74.12 | 1266.35±169.82 | 3918.05±155.11 |
| $T_2$ | 83.12±5.69 | 100.28±12.90 | 1064.94±214.37 |
| PD | 766.01±44.87 | 988.17±71.18 | 1737.31±55.78 |

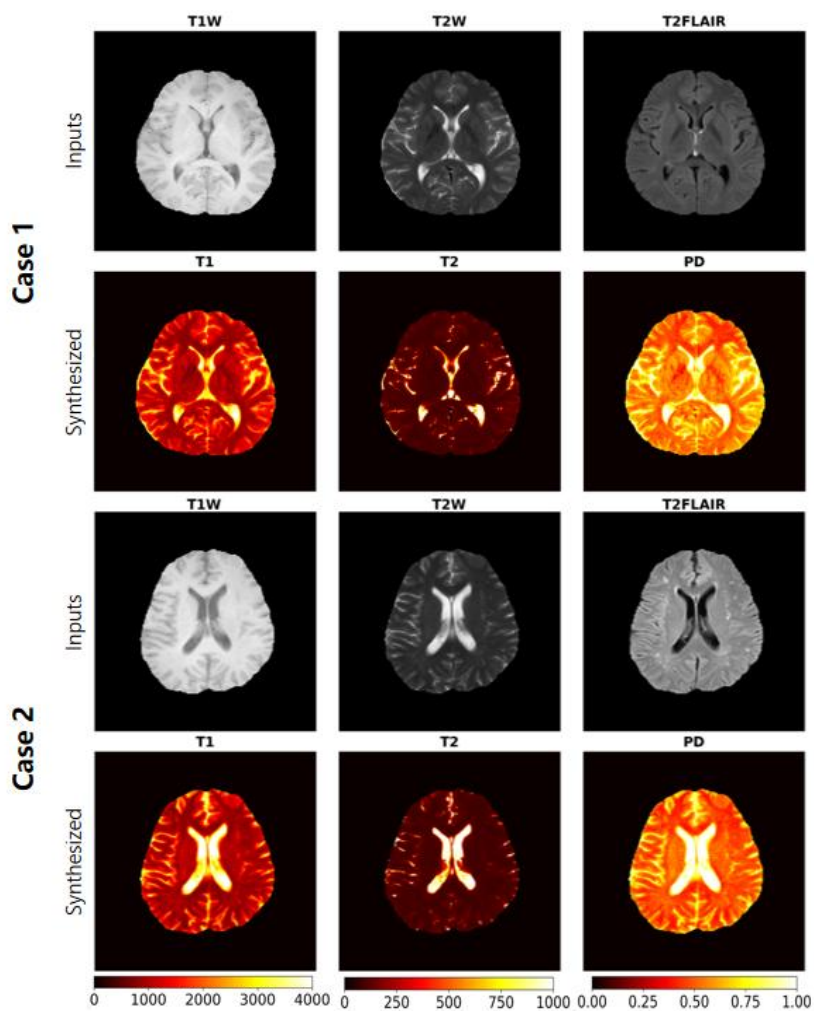

Figure 7. Representative axial slices of the $T_1$, $T_2$, and PD maps computed from external test set 2.

## 4. Discussion:

In this study, we proposed a physics-driven neural network for qMRI synthesis that explicitly addresses a critical limitation in current deep learning approaches - the integration of MRI physics principles into the learning framework and consideration of influences of scan parameters to the image contrast. We aim to learn the mapping from three primary contrast-weighted images ($T_1$w, $T_2$w, and $T_2$-FLAIR) to quantitative parametric maps ($T_1$, $T_2$, and PD).

We select $T_1$w, $T_2$w, and $T_2$-FLAIR images as the model input for two reasons. First, these three sequences are routinely scanned in a typical MRI examination. Secondly, there was a physical interdependence between these inputs and the target parametric maps. Each weighted image's signal intensity can be mathematically derived from the three parametric maps and the corresponding scan parameters. In addition, these contrasts provide complementary information about $T_1$ recovery, $T_2$ decay, and fluid, respectively.

In previous physics-guided approaches, Qiu et al.[29] estimate $T_1/T_2$ by feeding network-predicted maps back through Bloch-equation simulators to synthesize weighted images, training the network to minimize the difference between synthetic and actual acquisitions. Although this label-free strategy avoided the need for true $T_1/T_2$ ground-truth maps, it confines network applicability to images acquired with fixed scan parameters. Consequently, without explicit modeling of scan parameters' influence on image contrast, such approaches cannot fully capture the underlying MRI physics. In contrast, our proposed PDPE-Net explicitly includes the exact acquisition settings (TR, TE and TI) as part of the input and trains the network with a loss that more directly enforces quantitative accuracy. During inference, PDPE-Net generate quantitative maps from weighted images scanned with corresponding scan parameters, enabling flexible adaptation to varying acquisition conditions. As demonstrated in Figure 6, PDPE-Net can accurately generate quantitative maps even for previously unseen brain structures and lesions. This outcome highlights PDPE-Net's robustness and its capability to generalize effectively to novel and clinically relevant conditions, despite significant deviations in brain anatomy from the training data. Moreover, our approach shows strong potential as a data harmonization method, as it learns to map weighted images acquired under different scanning parameters into consistent and physically meaningful quantitative representations. This capability enables integration of heterogeneous MRI datasets and promotes more reliable downstream analysis in multi-center studies.

However, challenges remain in accurately modeling highly varying regions, such as CSF. While strong correlations were observed between predicted and true values in WM and GM, CSF predictions exhibited higher variability. There are several possible reasons. First, CSF $T_2$ error was expected because CSF signals were nulled on $T_2$-FLAIR, yielding low SNR for CSF $T_2$ information. Second, while parametric maps from MDME served as the reference standard, this technique was optimized for soft tissues with shorter $T_1/T_2$ values than CSF. Consequently, MDME-derived CSF reference values may be inaccurate which can be observed from Figure 2, which negatively influences the performance of PDPE-Net. We regard the suboptimal quantification of CSF values as a minor concern for clinical applications, given the limited diagnostic and prognostic relevance of CSF $T_1/T_2$ values. These findings suggest that further refinement of the parameter embedding strategies may be necessary to better model tissues with substantial inter-tissue differences, such as CSF, which poses particular challenges due to its low signal intensity and large variability across individuals. Furthermore, although our framework currently incorporates TE, TR, and TI parameters, extending it to include additional factors such as flip angle and magnetic field strength could further improve its generalizability and clinical adaptability. To this end, we are actively pursuing these directions and conducting broader evaluations across diverse cohorts and real-world clinical settings to validate the robustness and practical utility of our model under varying acquisition protocols.

**5. Conclusion**:

In this study, we proposed a physics-driven deep learning model with parameter-embedding for synthesizing quantitative MRI maps from weighted MRI images. The model can capture the underlying physics governing MRI signal intensities and can be used to generate parametric maps from weighted MRI images scanned with different scan parameters, making it ready for application in real-world clinical settings.

**Acknowledgments**

To maintain anonymity during the review process, acknowledgments are not included and will be provided in the final version if the paper is accepted.